\documentclass[]{interact}

\usepackage{epstopdf, multirow}
\usepackage[caption=false]{subfig}

\usepackage[numbers,sort&compress]{natbib}
\bibpunct[, ]{[}{]}{,}{n}{,}{,}
\renewcommand\bibfont{\fontsize{10}{12}\selectfont}
\makeatletter
\def\NAT@def@citea{\def\@citea{\NAT@separator}}
\makeatother

\theoremstyle{plain}

\theoremstyle{definition}

\theoremstyle{remark}

\begin{document}

\title{Incorporating Naive Bayes Classification to Address Subpopulation Structure in Familial DNA Search}

\author{
\name{Akaraphon Jantaraphum$^{1}$, Chanagarn Laoiam$^{1}$, Budsaba Rerkamnuaychoke$^{2}$, Jittima Shotivaranon$^{2}$, Monchai Kooakachai$^{1,*}$}
\affil{\textsuperscript{1}Department of Mathematics and Computer Science, Faculty of Science, Chulalongkorn University, Thailand}
\affil{\textsuperscript{2}Department of Pathology, Faculty of Medicine, Ramathibodi Hospital, Mahidol University, Thailand}
}

\maketitle

\begin{abstract}
Familial DNA search evaluates the genetic relatedness of two individuals by comparing the likelihood of their observed DNA profiles under two competing hypotheses—the null hypothesis that the individuals are unrelated and the alternative hypothesis that they are related—most commonly through the likelihood ratio (LR). Standard LR-based approaches typically assume a uniform genetic background; however, this assumption is rarely valid due to population substructure, where allele frequencies vary among subpopulations and can bias relationship inference. Existing modifications—such as LR calculations based on average allele frequencies (LRLAF) and strategies using maximum, minimum, or average likelihood ratios (LRMAX, LRMIN, LRAVG)—help mitigate these challenges but remain limited in their ability to fully address subpopulation differences. This study introduces a new LR-based statistic, LRCLASS, which incorporates a classification step using the Naive Bayes classifier to account for nuisance parameters associated with unknown subpopulation origins. In LRCLASS, the two DNA profiles being compared are jointly assigned to a subpopulation group via Naive Bayes before LR computation. Empirical evaluations using Thai population data show that LRCLASS achieves higher statistical power for detecting full-sibling relationships than existing LR-based methods. We further assessed multinomial logistic regression as an alternative classifier and found its performance comparable to that of Naive Bayes, suggesting flexibility in classifier choice. Overall, integrating the Naive Bayes classifier with LR computation offers a robust strategy for addressing population substructure in familial DNA search and highlights the broader potential of combining supervised learning techniques with forensic statistical methodologies to enhance the accuracy and reliability of genetic relationship testing.
\end{abstract}


\section{Introduction}
\subsection{Familial DNA Search}
In forensic investigations, genetic material or DNA is one of the most common types of evidence collected from crime scenes. It can be extracted from various parts of the human body, such as hair, blood, sweat, or saliva. A typical question in such cases is whether a DNA sample from the crime scene matches any individual in a forensic database. If a match is found, the investigation becomes more straightforward, as the individual who left the sample can be identified. The results can then be submitted to the court for further proceedings. However, if the DNA sample from the crime scene does not perfectly match anyone in the database but shows some similarities, it may indicate a genetic relationship between two individuals, such as parent-child or full-sibling. This situation leads to familial DNA searches, which use statistical tools to infer familial relationships based on DNA evidence. 

Various strategies for familial searching have been proposed in the literature, ranging from basic methods that count the number of shared alleles to more advanced probabilistic approaches (e.g., Balding \emph{et al.} 2013, Bieber \emph{et al.} 2006, Ge \emph{et al.} 2011, Ge and Budowle 2012, Kooakachai \emph{et al.} 2019, Kruijver \emph{et al.} 2014, Slooten and Meester 2014). Familial DNA searches have grown in prominence over the years. By 2014, at least 40 laboratories in the United States were conducting such searches, and approximately 75\% of those not yet practicing this method were actively considering its implementation (Debus-Sherrill and Field, 2019). Familial DNA search is one of several indirect methods used to identify potential sources of forensic biological evidence, along with approaches such as Y-STR database searches and investigative genetic genealogy, as noted by Ge and Budowle (2021). Although useful, this technique has raised ethical concerns, especially because nonwhite individuals are often overrepresented in forensic databases, increasing the likelihood that their families may be subjected to unnecessary or intrusive investigations. As a result, U.S. states have taken different regulatory approaches: some have created formal policies that either permit or prohibit familial DNA searches, while others allow investigations to proceed when partial matches arise accidentally, but only under specific oversight intended to reduce ethical risks (Katsanis, 2020).

In hypothesis testing for familial DNA searches, the likelihood ratio is frequently used to evaluate the probability of a pair of DNA profiles under two distinct relationship hypotheses. One scenario assumes that the individuals are unrelated, while the other considers a specific familial relationship, such as parent–child or full-sibling. Over the years, the adoption of the likelihood ratio statistic for relatedness testing has grown significantly. For example, a study on the New Zealand database showed that the likelihood ratio approach for parent–child and full-sibling tests outperformed the traditional method of counting matching alleles (Curran and Buckleton, 2008). Building on this, a validation study further demonstrated the effectiveness of a likelihood ratio–based method for identifying first-degree familial relationships using autosomal STR profiles in California’s 1,000,000-profile database (Myers \emph{et al.}, 2011). 

Although the likelihood ratio statistic is both straightforward and statistically robust, it has notable limitations. A key drawback is its assumption that all individuals in the population share the same genetic background, meaning their genetic information is drawn from a single allele frequency distribution. In forensic applications, however, this assumption is often unrealistic. Populations are more accurately modeled as having multiple genetic substructures, such as racial or ancestral groups, which can significantly influence allele frequencies. Failing to account for these differences can lead to inaccurate likelihood ratio calculations and erroneous conclusions, potentially resulting in miscarriages of justice, as suggested by Balding and Nichols (1994). In this context, population substructure refers to genetic variation that exists between distinct subgroups within a larger population. These variations must be incorporated into likelihood ratio calculations to ensure accuracy and fairness in forensic analyses.

The critical role of subpopulation structure in forensic science has therefore been emphasized in numerous studies to ensure accurate DNA analysis and minimize errors. Factors such as geographical proximity, migration patterns, and cultural practices shape subpopulation structures, directly influencing likelihood assessments in forensic investigations (Gallo \emph{et al.}, 1997). Analyses of allelic frequencies across racial and ethnic databases reveal substantial variations in probability calculations, and differences in allele frequencies—combined with genetic admixture—have been shown to increase the risk of false positives or reduce the statistical power of genetic studies (Tian \emph{et al.}, 2008). Strategies to address population substructure in DNA database searches have also been proposed; for instance, simulations using ten-locus DNA profiles show that even mild substructure can inflate evidence strength in 11.6\% of cases (Spooner and Stockmarr, 2019). A comprehensive overview of the importance of subpopulation structure has further been provided by Jeong \emph{et al.} (2021), who discuss foundational statistical concepts and recent advances, including methods for ancestral classification using informative single nucleotide polymorphism markers.

When allele frequencies are available for each subpopulation, several adjustments to the likelihood ratio formula have been proposed. One notable approach, implemented by the Denver Crime Laboratory in Colorado, USA, in 2008, was inspired by the work of Prescutti \emph{et al.} (2002). This method involves calculating likelihood ratios multiple times, applying a single-population framework to each available allele frequency distribution. The highest value from these calculations, referred to as LRMAX, is then selected as the final likelihood ratio statistic. Alternatively, Ge and Budowle (2012) proposed using the minimum likelihood ratio instead, arguing that it provides a more accurate kinship index for the population associated with the forensic profile. Their research demonstrated that the minimum likelihood ratio consistently outperformed the maximum method in terms of accuracy. Two additional likelihood ratio–based statistics were introduced in 2019: LRLAF, which applies the classical likelihood ratio framework using average available allele frequencies, and LRAVG, which calculates the average of all likelihood ratios derived from each allele frequency distribution (Kooakachai \emph{et al.}, 2019). When these methods were compared for parent–child and full-sibling tests in the Colorado population, LRLAF showed approximately 10\% higher power for parent–child tests and 5\% higher power for full-sibling tests compared to LRMAX. However, debate continues regarding the consistency and generalizability of these methods, particularly when applied to populations outside of Colorado.

Existing methods for familial DNA searches in the literature primarily rely on basic arithmetic and do not incorporate advanced classification techniques. Rather than simply averaging likelihood ratios (LRAVG), averaging at the allelic level (LRLAF), or using the maximum (LRMAX) or minimum (LRMIN), machine learning approaches can be used to predict the subpopulation to which a DNA profile belongs, followed by likelihood ratio calculation after classification. The use of classification has already been applied in certain forensic subfields. For example, penalized logistic regression was employed in forensic toxicology to calculate likelihood ratios in a two-class classification setting (Biosa \emph{et al.}, 2019). The effectiveness of this framework was demonstrated through a case study on alcohol biomarker data, highlighting its ability to distinguish chronic alcohol drinkers and its potential adaptability for broader forensic and analytical chemistry applications.

This study introduces a novel likelihood ratio–based statistic designed to evaluate the genetic relationship between two individuals while explicitly accounting for population substructure. The approach begins by classifying each DNA profile into a subpopulation using a Naive Bayes classifier, after which the likelihood ratio is computed based on the predicted subpopulations. This framework aims to improve the accuracy and robustness of familial inference compared with existing LR-based approaches. In addition, we implement an alternative classification method using multinomial logistic regression, enabling a systematic comparison of classification performance. Section 2 describes how DNA profiles are stored and presents both existing and proposed statistics. Section 3 outlines the algorithms used to evaluate statistical power and compare classification strategies. Section 4 presents the results, and Sections 5 and 6 provide discussion and conclusions.

\section{Methods} \label{sec2}

\subsection{DNA Profile and Dataset Description}

Humans differ due to the random combination of nucleobases—adenine (A), thymine (T), guanine (G), and cytosine (C)—inherited from both parents. The human genome, composed of about 3 billion nucleobases and approximately 20,000 genes across 23 pairs of chromosomes, contains all the genetic instructions for an individual. Genetic markers are variations in DNA sequences at specific positions on chromosomes. Among these markers, short tandem repeats (STRs) are especially useful in forensic science. STRs consist of DNA sequences, typically two to seven base pairs long, repeated multiple times. The number of repeats at each locus varies between individuals, allowing for identification. For example, a 28-base pair sequence like TCTATCTATCTATCTATCTATCTATCTA represents 7 repeats of the four-base sequence TCTA, indicating an allele of length 7 at that position. Each person inherits one STR allele from each biological parent, and the number of repeats can differ between them. A genotype at a specific locus is represented by two alleles, such as 11,12, meaning the individual has 11 and 12 repeats of the TCTA sequence at that locus. The variation in STR repeat lengths among individuals makes STRs highly effective for forensic identification.

In this paper, a DNA profile is defined as the collection of genotypes across multiple loci for an individual, with the number of loci used in forensic investigations varying by region. Thailand is highlighted as a case study due to its well-documented genetic substructure and diversity, supported by archaeological and linguistic research. Genetic differences have been observed between Thai-Malay Muslims, primarily in southern Thailand, and Thai-Buddhists from other regions (Kutanan \emph{et al.}, 2014a). Distinct genetic profiles have also been identified between northeastern Thai populations and other groups, including those in northern Thailand, through mitochondrial DNA HVR1 variation studies conducted in the same year (Kutanan \emph{et al.}, 2014b). More recently, genome-wide analyses of 32 Thai subpopulations have further emphasized significant genetic differences, highlighting the importance of careful participant selection in biomedical and clinical research (Kutanan \emph{et al.}, 2021).

Although Thailand does not yet have a National DNA Database, a DNA database was established by the Central Institute of Forensic Science (CIFS) in 2004 to address crimes in the Deep South. This database has since expanded to include approximately 160,000 profiles from crime samples, individuals, and prisoners, utilizing both autosomal and Y-STR markers. In 2016, a 24-loci STR kit was introduced to support both direct and familial matching (Boonderm \emph{et al.}, 2017). DNA databases, including government repositories, play a crucial role in criminal investigations by helping identify potential suspects. STR markers are particularly important because of their high variability and strong discriminatory power, which makes them essential for precise identification (Butler, 2005).

To perform analyses relevant to Thailand’s genetic diversity, this study focuses on four major Thai subpopulations defined by geographical regions: North (NO), Northeastern (NE), Central (CT), and South (SO). Allele frequency distributions for these groups were obtained from DNA samples of 929 unrelated Thai individuals, analyzed across 15 loci (Shotivaranon \emph{et al.}, 2009). The sample sizes for the North, Northeastern, Central, and South subpopulations were 202, 304, 212, and 211, respectively. Based on the available data, the 15 core loci used in this study are: D8S1179, D21S11, D7S820, CSF1P0, D3S1358, TH01, D13S317, D16S539, D2S1338, D19S433, vWA, TPOX, D18S51, D5S818, and FGA.

\subsection{Hypothesis Test}

The task of determining whether two DNA profiles $X_1$ and $X_2$ share a specific genetic relationship can be framed as a hypothesis test: 
\begin{align*}
    H_0: & \text{ The two individuals are unrelated.} \\
    H_1: & \text{ The two individuals share a specific genetic relationship.}
\end{align*}

The genetic relatedness between two individuals can be described using the notation proposed by Lynch and Walsh (1998). They suggested that all familial relationships can be explained through the probabilities of alleles at a locus being identical by descent (IBD), where two alleles are considered IBD if they are inherited as copies of the same ancestral allele. For any two individuals, the relatedness at a specific locus is expressed as a 3-tuple {\boldmath$\theta$} = $(z_{0}, z_{1}, z_{2}).$ 
Here, $z_0$ represents the probability that neither allele of $X_1$ is IBD to any allele of $X_2$, $z_1$ refers to the probability that exactly one allele of $X_1$ is IBD to an allele of $X_2,$ and $z_2$ indicates the probability that both alleles of $X_1$ are IBD to those of $X_2$. Using this definition, the parent-child relationship is represented by {\boldmath$\theta$} = (0,1,0), as the parent and child always share exactly one ancestral allele at each locus. Similarly, the unrelated and full-sibling relationships are expressed as {\boldmath$\theta$} = (1,0,0) and (1/4, 1/2, 1/4), respectively. It is important to note that, by definition, the probabilities $z_{0}, z_{1},$ and $z_{2}$ must sum to one. 

The hypotheses for testing a parent-child relationship can be rewritten as the following:
\begin{center}
$H_{0}$ : {\boldmath$\theta$} = (1,0,0) \,\, versus \,\, $H_{1}$ : {\boldmath$\theta$} = (0,1,0), 
\end{center}
whereas testing for a full-sibling relationship corresponds to 
\begin{center}
$H_{0}$ : {\boldmath$\theta$} = (1,0,0) \,\, versus \,\, $H_{1}$ : {\boldmath$\theta$} = ($\frac{1}{4},\frac{1}{2},\frac{1}{4}$). 
\end{center}
The test statistic used is the likelihood ratio $\Lambda$ defined by
\begin{align*}
\Lambda
&= \dfrac{P(X_1, X_2 \mid H_{1})}{P(X_1, X_2 \mid H_{0})}. 
\end{align*}
If $\Lambda > 1$, then there is more chance that the two DNA profiles $X_1$ and $X_2$ were generated from $H_1$ compared to $H_0$. In particular, higher $\Lambda$ corresponds to more evidence to reject the null hypothesis and conclude a specific relationship of interest. 

Calculating the probabilities in the numerator and denominator of $\Lambda$ requires knowledge of the allele frequency distributions. Let $f_{X_1}$ and $f_{X_2}$ represent the allele frequency distributions for the populations to which $X_1$ and $X_2$ belong, respectively. These distributions provide the frequencies of each allele, allowing the probabilities to be fully determined. In other words, $f_{X_1}$ and $f_{X_2}$ are essential inputs for calculating the likelihood ratio, serving as nuisance parameters. Additionally, the probabilities are calculated separately for each locus and then combined by multiplying them, based on the assumption that loci are independent. The assumption of independence between loci requires that the loci are not genetically linked and are in linkage equilibrium. In fact, some core loci do not fully satisfy this assumption. For instance, the loci vWA and D12S391 are located on the same chromosome, approximately 6.3 megabases (Mb) apart. Ideally, genetic markers on the same chromosome should be separated by more than 50 Mb to ensure independence between genotypes at the two loci (O'Connor \emph{et al.}, 2011). Studies have indicated that independence between vWA and D12S391 cannot always be assumed at the individual level. Nevertheless, at the population level, there is no significant evidence of correlation between genotypes at the core loci, including these two loci (Budowle \emph{et al.} 2011, Ge \emph{et al.} 2012). For simplicity in our calculations, we proceed under the assumption of independence between loci.

In the simplest case of a population with no substructure, $f_{X_1}$ and $f_{X_2}$ are equal and correspond to the allele frequency distribution of the entire population, making the likelihood ratio test well-defined. However, when the population exhibits substructure, $f_{X_1}$ and $f_{X_2}$ can take on multiple possible values. Currently, there is no established theory identifying the optimal test statistic for such cases. This poses a significant challenge in determining the most powerful statistic for hypothesis testing in populations with substructure.

\subsection{Existing Test Statistics}

Consider a population divided into $R > 1$ distinct subpopulations, each with its own allele frequency distribution, denoted as $f_i$ for the $i^{\text{th}}$ subpopulation. We define four existing methods for conducting familial DNA testing when the profiles $X_1$ and $X_2$ are given.

The first statistic is obtained by using the likelihood ratio framework but plugging in the value of $f_{X_1}$ and $f_{X_2}$ by $f_{local}$ which is defined by 
$$f_{local }= p_1f_1+p_{2}f_{2}+ \cdots +p_{R}f_{R}$$
where $p_i$ is the proportion of individuals in the $i^{\text{th}}$ subpopulation. This statistic is called LRLAF and can be written mathematically as 
$$ \Lambda_{LAF} = \frac{L(\theta_1 \mid X_1, X_2, f_{local})}{L(\theta_0 \mid X_1, X_2, f_{local})}. $$

Examining the LRLAF approach reveals that it involves averaging directly at the allele frequency level. A slight modification can be made by shifting the averaging process to the likelihood ratio level instead. This leads to a new statistic called LRAVG. In this method, the likelihood ratios are computed repeatedly using the single-population framework applied to multiple locally available allele frequency distributions. The final step involves calculating a weighted average of these likelihood ratios. Mathematically, the LRAVG statistic is expressed as follows:

$$ \Lambda_{AVG} = p_1\frac{L(\theta_1 \mid X_1, X_2, f_1)}{L(\theta_0 \mid X_1, X_2, f_{1})}+p_2\frac{L(\theta_1 \mid X_1, X_2, f_2)}{L(\theta_0 \mid X_1, X_2, f_2)}+ \dots + p_R\frac{L(\theta_1 \mid X_1, X_2, f_R)}{L(\theta_0 \mid X_1, X_2, f_R)}. $$

If instead of calculating a weighted average in the final step, we opt to take the maximum or minimum, the resulting statistics are referred to as LRMAX or LRMIN, respectively. The definitions of these two statistics are as follows:

$$ \Lambda_{MAX} =\max\Bigl\{\frac{L(\theta_1 \mid X_1, X_2, f_1)}{L(\theta_0 \mid X_1, X_2, f_1)},\frac{L(\theta_1 \mid X_1, X_2, f_2)}{L(\theta_0 \mid X_1, X_2, f_2)}, \dots ,\frac{L(\theta_1 \mid X_1, X_2, f_R)}{L(\theta_0 \mid X_1, X_2, f_R)}\Bigr\}. $$

$$ \Lambda_{MIN} =\min\Bigl\{\frac{L(\theta_1 \mid X_1, X_2, f_1)}{L(\theta_0 \mid X_1, X_2, f_1)},\frac{L(\theta_1 \mid X_1, X_2, f_2)}{L(\theta_0 \mid X_1, X_2, f_2)}, \dots ,\frac{L(\theta_1 \mid X_1, X_2, f_R)}{L(\theta_0 \mid X_1, X_2, f_R)}\Bigr\}. $$

\subsection{Proposed Test Statistic}

The key to successfully computing the likelihood ratio lies in eliminating the nuisance parameters $f_{X_1}$ and $f_{X_2}$. Existing methods typically rely on basic statistics like the average, maximum, and minimum. However, more advanced statistical techniques can be employed to predict the true values of $f_{X_1}$ and $f_{X_2}$. An example of this can be seen in the application of Naive Bayes classification.

Recall that we have a population divided into $R > 1$ distinct subpopulations. Given a DNA profile $B$, the task is to determine which subpopulation it belongs to. Let $A_i$ be the event such that a DNA profile $B$ corresponds to the $i^{\text{th}}$ subpopulation. By using the Bayes' rule, for $i\ \in\{1,\ 2,\ \ldots,\ R\},$ we have 
\[
P\left(A_i\mid B\right)=\frac{P\left(B\mid A_i\right)P\left(A_i\right)}{\sum_{k=1}^{R}{P\left(B\mid A_k\right)P\left(A_k\right)}}\ .  \label{bayes}
\]

On the right-hand side of the equation, the subpopulation prior probability $P\left(A_i\right)$ can be determined using the proportion of each subpopulation. The term $P\left(B \mid A_i\right),$ which represents the probability of observing a DNA profile $B$ given that the associated subpopulation is $i,$ can be computed using allele frequency data from the $i^{\text{th}}$ subpopulation. An illustrative example is presented here: consider a scenario with $R = 4$ subpopulations and three independent loci, namely TPOX, vWA, and D13S317. The DNA profiles of two individuals, $X_1$ and $X_2$,  are presented in Table \ref{tab: ex DNA}. 

\begin{table}[t]
\footnotesize
\centering
\caption{An example of DNA profiles}\label{tab: ex DNA}
\tabcolsep=35.5pt
\begin{tabular}{cccc}
\hline \hline
\multirow{2}*{DNA profiles} & \multicolumn{3}{c}{Locus} \\\cline{2-4}
         & TPOX & vWA & D13S317 \\\hline
$X_1$  & $(10,10)$ & $(15,17)$ & $(9,10)$ \\
$X_2$  & $(10,11)$ & $(15,15)$ & $(9,10)$ \\
\hline \hline
\end{tabular}
\end{table}

Additionally, assume that the allele frequency distributions for the specified loci are provided in Table \ref{tab: ex al freq}. Assume further that the prior probabilities for each subpopulation are $P(A_1) = 0.1, P(A_2) = 0.2, P(A_3) = 0.3$ and $P(A_4) = 0.4.$ 

\begin{table}[t]
\footnotesize
\centering
\caption{The allele frequency distributions for the specified loci with respect to the example.} \label{tab: ex al freq}
\tabcolsep=14.5pt
\begin{tabular}{cccccc}
\hline \hline
Allele & Locus & prob. subpop1 & prob. subpop2 & prob. subpop3 & prob. subpop4 \\\hline
10 & TPOX & 0.0223 & 0.0378 & 0.0354 & 0.0569 \\
11 & TPOX & 0.2351 & 0.2072 & 0.2618 & 0.2867 \\
12 & TPOX & 0.0173 & 0.0263 & 0.0071 & 0.0237 \\
\hline
15 & vWA & 0.0396 & 0.0263 & 0.0283 & 0.0332 \\
16 & vWA & 0.1015 & 0.0872 & 0.1297 & 0.1445 \\
17 & vWA & 0.2475 & 0.2484 & 0.2524 & 0.2725 \\
\hline
9 & D13S317 & 0.1386 & 0.1464 & 0.1344 & 0.1398 \\
10 & D13S317 & 0.1609 & 0.1168 & 0.1179 & 0.0995 \\
11 & D13S317 & 0.2327 & 0.2072 & 0.2406 & 0.2512 \\
\hline \hline
\end{tabular}
\end{table}

We consider the simultaneous classification of $X_1$ and $X_2$. Given the independence of the two profiles, we have:
$$P\left(X_1, X_2 \mid A_i\right)=P\left(X_1 \mid A_i\right)P\left(X_2 \mid A_i\right).$$
To calculate the genotype probability across $m$ loci, we assume that these loci are independent. Therefore, the genotype probability across $m$ loci is the product of the genotype probabilities at each locus. Specifically, for $k = 1,2$, we have
$$
P\left(X_k \mid A_i\right)=\prod_{j=1}^{m}{P\left(X_{k,j}\mid A_i\right)} $$
where $X_{k,j}$ represents the genotype of $X_k$ at locus $j$.  It is important to note that the genotype at locus $j$ can be of two types: homozygous (aa) or heterozygous (ab). The corresponding probabilities for these genotypes are $p^2$ and $2pq,$ where $p$ and $q$ are the allele frequencies of alleles a and b at locus $j$, respectively. In the context of this example, we set $m = 3$, and an example of the complete calculation of genotype probabilities for the case where $k =1$ and $i=1$ is shown as follows:
\begin{align*}
P\left(X_1 \mid A_1 \right) &= \prod_{j=1}^{3}{P\left(X_{1,j}\mid A_1 \right)} = (0.0223^2)(2(0.0396)(0.2475))(2(0.1386)(0.1609)) \\ &=4.3477 \times 10^{-7}
\end{align*}

The calculations for other values of $k$ and $i$ can be performed in the same manner. When all values are combined, the results in this context are as follows: $P\left(A_1 \mid X_1,X_2 \right) = 0.2903, P\left(A_2 \mid X_1,X_2 \right) = 0.0431, P\left(A_3 \mid X_1,X_2 \right) = 0.0729,$ and $P\left(A_4 \mid X_1,X_2 \right) = 0.5938$.  
 
In general, after calculating $P\left(A_i\mid B\right)$ for all $i\ \in\{1,\ 2,\ldots,\ R\}$, the subpopulation with the highest probability is identified as the predicted origin of the DNA profile $B$. In other words, the DNA profile $B$ is classified as:
\[
\arg\max_i  \{P\left(A_i\mid B\right)\} .\]
Therefore, in this specific example, DNA profile $X_1$ and $X_2$ are classified into the $4^{\text{th}}$ subpopulation, as it has the highest probability. 

Next, we formally define the proposed statistic. Given the DNA profiles of two individuals, we first use a Naive Bayes classifier to jointly determine their most likely subpopulation. Once the classification is obtained, the likelihood ratio is computed under a single-population framework corresponding to the assigned subpopulation. This statistic, denoted as LRCLASS, is mathematically expressed as:
$$ \Lambda_{CLASS}:=\frac{L(\theta_1 \mid X_1, X_2, f_{result})}{L(\theta_0 \mid X_1, X_2, f_{result})} $$
where $f_{result}$ represents the subpopulation predicted from the classification.

\section{Simulation Study} \label{sec2}

\subsection{Algorithms for Power Comparison}

To compare the existing statistics with the proposed statistic, we evaluate them based on statistical power. The power of each statistic will be determined through simulations. These simulations will involve a sample size of 1,000,000 pairs under the null hypothesis and 1,000,000 pairs under the alternative hypothesis. The numerical analysis will be conducted using the Thai population as a case study, focusing on parent-child and full-sibling relationships.

Recall that power refers to the probability of correctly rejecting the null hypothesis when the alternative hypothesis is true. To determine this, we first generate a null distribution to establish a criterion for rejecting the null hypothesis, and then create the alternative hypothesis to estimate the power. The algorithm is outlined below.

\begin{enumerate}
\item Simulate a pair of unrelated DNA profiles using $m=15$ loci. The subpopulation information of the individuals is randomly selected from a multinomial distribution with success probabilities $p_{NO} = 0.11083, \, p_{NE} = 0.36944, \, p_{CT} = 0.35383, \, p_{SO} = 0.16590$. These estimates are derived from the Thai population database provided by the National Statistical Office of Thailand (2023).

\item	Calculate the likelihood ratio statistics for the pair drawn in Step 1 using five different formulas: LRLAF, LRAVG, LRMIN, LRMAX, and LRCLASS. Save the resulting values.
\item Repeat Steps 1 and 2 many times, for example, $B =$ 1,000,000 times. This will generate the null distribution of likelihood ratios.
\item	Estimate a threshold $c$ from the null distribution obtained in Step 3. Specifically, we determine $c$ such that $P(\text{LR} > c) = \alpha$, where $\alpha$ is the specified false positive rate.
\item Generate a pair of DNA profiles based on the relationship in the alternative hypothesis (using $m = 15$ loci). For each pair, the subpopulation information of the first individual is selected randomly from a multinomial distribution, as introduced in Step 1, while the subpopulation information of the second individual is forced to match the first, as they are genealogically related.
\item	Calculate the likelihood ratio statistics for the pair drawn in Step 5 using five different formulas: LRLAF, LRAVG, LRMIN, LRMAX, and LRCLASS. Save the resulting values.
\item	Repeat Steps 5 and 6 multiple times (for instance, 1,000,000 times) to obtain the alternative distribution of likelihood ratios based on the generated samples.
\item Estimate the power by calculating the ratio of the number of pairs with statistics greater than the threshold $c$ to the total number of simulated pairs in Step 7.
\end{enumerate}

Once the power for each statistic is obtained, we aim to compare them at the same false positive rate (FPR) $\alpha$. To determine an appropriate value for $\alpha$, we can refer to previous studies in the Colorado prison population, which suggested that for parent-child and full-sibling tests, $\alpha$ should be $1.7 \times 10^{-5}$ and $1.2 \times 10^{-5},$ respectively (Kooakachai \emph{et al.}, 2019).

\subsection{Classification Comparison}

We have developed the LRCLASS statistic, which classifies two DNA profiles into a single subpopulation simultaneously. The classification method employed was the Naive Bayes classifier. However, in statistics and machine learning, numerous other classification techniques are available. In this section, we describe an alternative classification method—multinomial logistic regression—and discuss how its performance compares to the previously proposed Naive Bayes classifier.

To begin, two approaches are outlined for representing DNA profiles as explanatory variables in logistic regression. A DNA profile with $m$ loci can be expressed in the general format:
$$\{ (x_{1a},x_{1b}),(x_{2a},x_{2b}),(x_{3a},x_{3b}),\ldots,(x_{ma},x_{mb})\}$$
where $x_{ij}$ denotes the allele observed at locus $i \in \{1, 2, \ldots, m\}$ and $j \in \{a, b\}$ is the maternal or paternal origin of the allele. The target variable in the proposed logistic regression is the subpopulation to which an individual belongs. Given the consideration of multiple subpopulations, a multinomial logistic regression model is utilized. The predictors can be structured in two distinct ways:
\begin{description}
\item{\underline{Method A}} Treat each allele count $x_{ij}$ as a categorical variable and use them directly as predictors in the regression model. This results in $2m$ categorical variables.
\item{\underline{Method B}} Represent the genotype at each locus as a single categorical variable. Specifically, define $u_i = (x_{ia}, x_{ib})$ as the predictor for locus $i$. This approach reduces the number of predictors to $m$, but each variable includes a larger number of categorical levels compared to Method A.
\end{description}

Logistic regression will yield the probabilities of each DNA profile belonging to each subpopulation. The predicted subpopulation is then determined as the one with the highest probability. 

To evaluate the performance of three classification methods—logistic regression with Method A, logistic regression with Method B, and Naive Bayes classification—we employ the following algorithm:
\begin{enumerate}
\item A total of 100,000 Thai DNA profiles are generated based on allele frequency distributions reported by Shotivaranon \emph{et al.} (2009).
\item The dataset from Step 1 is divided into five equal parts. For each trial, four parts (80\% of the dataset) are used as the training set, while the remaining part (20\%) is designated as the validation set. This process is repeated five times, ensuring that each part serves as the validation set exactly once. This approach is analogous to the concept of 5-fold cross-validation.
\item In each trial, subpopulation prediction is performed on the validation set using three distinct methods: logistic regression with Method A (requiring the training set), logistic regression with Method B (requiring the training set), and Naive Bayes classification (not requiring the training set). The corresponding validation errors are recorded in the form of a confusion matrix and summarized using standard metrics such as precision, recall, specificity, accuracy, error, and F1-score, which will be further discussed in the subsequent section.
\item The analyses in Step 3 are conducted five times, each based on a different validation dataset, and the average of these five repetitions will be used for comparison.
\item We also utilize the test dataset from Shotivaranon \emph{et al.} (2009), which includes 929 samples from Thai individuals. Subpopulation prediction is performed on this test dataset using three different methods: logistic regression with Methods A and B (both requiring the 100,000 generated samples from Step 1 as the training data), and Naive Bayes classification (which does not require a training set). The corresponding test errors are then recorded.
\item The performance of the three classification methods is compared by examining the test errors obtained in Step 5. 
\end{enumerate}

\subsection{Evaluation Metrics for Multi-class Classification}

We begin by reviewing four standard classification terms. A True Positive (TP) is a sample correctly identified as positive, while a True Negative (TN) is correctly identified as negative. A False Positive (FP) occurs when a negative sample is incorrectly labeled as positive, and a False Negative (FN) occurs when a positive sample is incorrectly labeled as negative. The corresponding classification metrics are defined as follows:
\begin{itemize}
    \item \textbf{Accuracy}: the number of samples correctly classified out of all samples in the test set, calculated as 
    \[
        \text{Accuracy} = \frac{\text{TP} + \text{TN}}{\text{TP} + \text{TN} + \text{FP} + \text{FN}} \in [0,1].
    \]
    \item \textbf{Precision}: the number of samples correctly classified as positive out of all samples predicted to be positive by the model. Mathematically, it is 
    \[
        \text{Precision} = \frac{\text{TP}}{\text{TP} + \text{FP}} \in [0,1].
    \]
    \item \textbf{Recall}: the number of samples correctly predicted to belong to the positive class out of all samples that actually belong to the positive class, written as 
    \[
        \text{Recall} = \frac{\text{TP}}{\text{TP} + \text{FN}} \in [0,1].
    \]
    \item \textbf{Specificity}: the number of samples correctly predicted to be in the negative class out of all samples that actually belong to the negative class, given by 
    \[
        \text{Specificity} = \frac{\text{TN}}{\text{TN} + \text{FP}} \in [0,1].
    \]
    \item \textbf{Error}: the number of samples incorrectly classified out of all samples in the test set, given by 
    \[
        \text{Error} = \frac{\text{FP} + \text{FN}}{\text{TP} + \text{TN} + \text{FP} + \text{FN}} \in [0,1].
    \]
    \item \textbf{F1-score}: the harmonic mean of precision and recall for the positive class, defined as 
    \[
        \text{F1-score} = \frac{2 \times \text{Precision} \times \text{Recall}}{\text{Precision} + \text{Recall}} \in [0,1].
    \]
\end{itemize}

For a classification task with $n$ samples distributed among $R \ge 3$ classes, the classifier’s performance can be summarized using an $R \times R$ confusion matrix. Each entry $n_{ij}$ at the intersection of row $i$ and column $j$, for $i,j = 1,\ldots,R$, represents the number of samples truly belonging to class $i$ but predicted as class $j$. The same evaluation metrics used in binary classification can be extended to this multiclass setting.

Two common strategies can be used to compute evaluation metrics (Rainio \emph{et al.}, 2024). The first constructs a separate $2 \times 2$ confusion matrix for each of the $R$ classes, with components defined as:
\[
    \mathrm{TP}_i = n_{ii}, \quad 
    \mathrm{TN}_i = \sum_{j \ne i} \sum_{k \ne i} n_{jk}, \quad
    \mathrm{FN}_i = \sum_{j \ne i} n_{ij}, \quad
    \mathrm{FP}_i = \sum_{j \ne i} n_{ji},
\]
for $i = 1,\ldots,R$. In the macro-averaging approach, each metric is first computed separately for each class using $\mathrm{TP}_i$, $\mathrm{TN}_i$, $\mathrm{FN}_i$, and $\mathrm{FP}_i$, and the final score is obtained by averaging the class-specific results. In contrast, micro-averaging aggregates the totals across all classes—$\sum_{i=1}^R \mathrm{TP}_i$, $\sum_{i=1}^R \mathrm{TN}_i$, $\sum_{i=1}^R \mathrm{FN}_i$, and $\sum_{i=1}^R \mathrm{FP}_i$—and computes each metric from these combined counts. The essential difference is that macro-averaging gives equal weight to each class, while micro-averaging weights each individual sample equally, thus making it more sensitive to class imbalance.

The next two metrics are Cohen’s Kappa $(\kappa)$ and the Matthews Correlation Coefficient (MCC), as presented in Rainio \emph{et al.} (2024). Both measures offer advantages over basic evaluation metrics in multi-class settings, particularly when dealing with imbalanced class distributions. Cohen’s Kappa incorporates an adjustment for agreement occurring by chance, making it more dependable when class proportions differ substantially. MCC, on the other hand, accounts for all entries in the confusion matrix and yields a more balanced, comprehensive assessment of classifier performance. Their formulas are given as follows:

\begin{align*}
    \kappa &= \frac{p_0 - p_e}{1 - p_e} \quad\text{with}\quad p_0 = \frac{1}{n} \sum_{i=1}^R n_{ii}, \quad p_e = \frac{1}{n^2} \sum_{i=1}^R n_{i\cdot}n_{\cdot i}, \\
    \text{MCC} &= \frac{n \sum_{i=1}^R n_{ii} - \sum_{i=1}^R n_{i\cdot}n_{\cdot i}}{\sqrt{(n^2 - \sum_{i=1}^R n_{i\cdot}^2)(n^2 - \sum_{i=1}^R n_{\cdot i}^2)}},
\end{align*}
where $n_{i\cdot} = \sum_{j=1}^R n_{ij}$, $n_{\cdot i} = \sum_{j=1}^R n_{ji}$ and $n = \sum_{i=1}^R \sum_{j=1}^R n_{ij}$. We will evaluate and compare the classification performance using the metrics outlined above.

\section{Results}

\subsection{Power Comparison}
Table \ref{threstable} displays the critical thresholds for rejecting the null hypothesis associated with each test statistic, thereby defining the rejection regions for the corresponding hypothesis tests. For example, when using the LRCLASS statistic to assess the parent-child relationship test at a FPR of $1.7 \times 10^{-5}$, the suggested rejection region consists of values equal to or exceeding 6860. 

\begin{table}[t]
\footnotesize
\centering
\caption{Parent-child (PC) and full-sibling (SB) test thresholds}\label{threstable}
\tabcolsep=27.5pt
\begin{tabular}{ccc}
\hline \hline
Test statistic & PC threshold at $\alpha = 1.7 \times 10^{-5}$ & SB threshold at $\alpha = 1.2 \times 10^{-5}$ \\\hline
LRLAF    & 8251 & 10280 \\ 
LRAVG    & 25111 & 28429 \\ 
LRMAX    & 46070 & 79538 \\ 
LRMIN    & 5666 & 5835 \\ 
LRCLASS  & 6860 & 5835 \\ 
\hline \hline
\end{tabular}
\end{table}

Subsequently, Table \ref{powertable} shows the power of each test for the parent-child relationship at an FPR of $1.7 \times  10^{-5}$ and for the full-sibling relationship at an FPR of $1.2 \times 10^{-5}$. Additionally, the table includes 95\% confidence intervals for the power estimates, calculated using the exact Binomial test, to provide a robust measure of precision.

\begin{table}[t]
\footnotesize
\centering
\caption{Power for the parent-child (PC) test at an FPR of $1.7 \times  10^{-5}$ and for the full-sibling (SB) test at an FPR of $1.2 \times 10^{-5}$} \label{powertable}
\tabcolsep=18pt
\begin{tabular}{ccccc}
\hline \hline
Statistic & PC Power & 95\% CI & SB Power & 95\% CI \\\hline
LRLAF    & 0.8383 & (0.83754, 0.83898) & 0.5431 & (0.54215, 0.54410) \\
LRAVG    & 0.7773 & (0.77647, 0.77810) & 0.5110 & (0.51003, 0.51199) \\ 
LRMAX    & 0.7814 & (0.78062, 0.78224) & 0.4885 & (0.48751, 0.48947) \\ 
LRMIN    & 0.8176 & (0.81688, 0.81840) & 0.5318 & (0.53085, 0.53280) \\
LRCLASS  & 0.8116 & (0.81079, 0.81233) & 0.5748 & (0.57388, 0.57582) \\ 
\hline \hline
\end{tabular}
\end{table}

For the parent-child relationship test, as presented in Table \ref{powertable}, the LRLAF statistic demonstrates the highest power, achieving approximately 83.8\%. This is followed by LRMIN and LRCLASS, both exhibiting power values in the range of 81–82\%. In contrast, LRMAX and LRAVG yield comparatively lower power, approximately 78\%.

The primary objective of this study, however, is to assess the performance of the proposed statistic, LRCLASS. In the context of parent-child relationship testing, LRCLASS ranks third, with a power that is 2.7\% lower than the highest-ranking LRLAF and 0.6\% lower than the second-ranking LRMIN. In the full-sibling relationship test, LRCLASS exhibits superior performance, achieving the highest power among the five evaluated statistics at a FPR of $1.2 \times 10^{-5}$. Notably, LRCLASS surpasses the second-ranking statistic, LRLAF, by a margin of approximately 3.2\%. The ranking of the statistics for the full-sibling case, from highest to lowest performance, is as follows: LRCLASS, LRLAF, LRMIN, LRAVG, and LRMAX.

\begin{figure}[t]
\centering
\includegraphics[height=12cm,keepaspectratio=true]{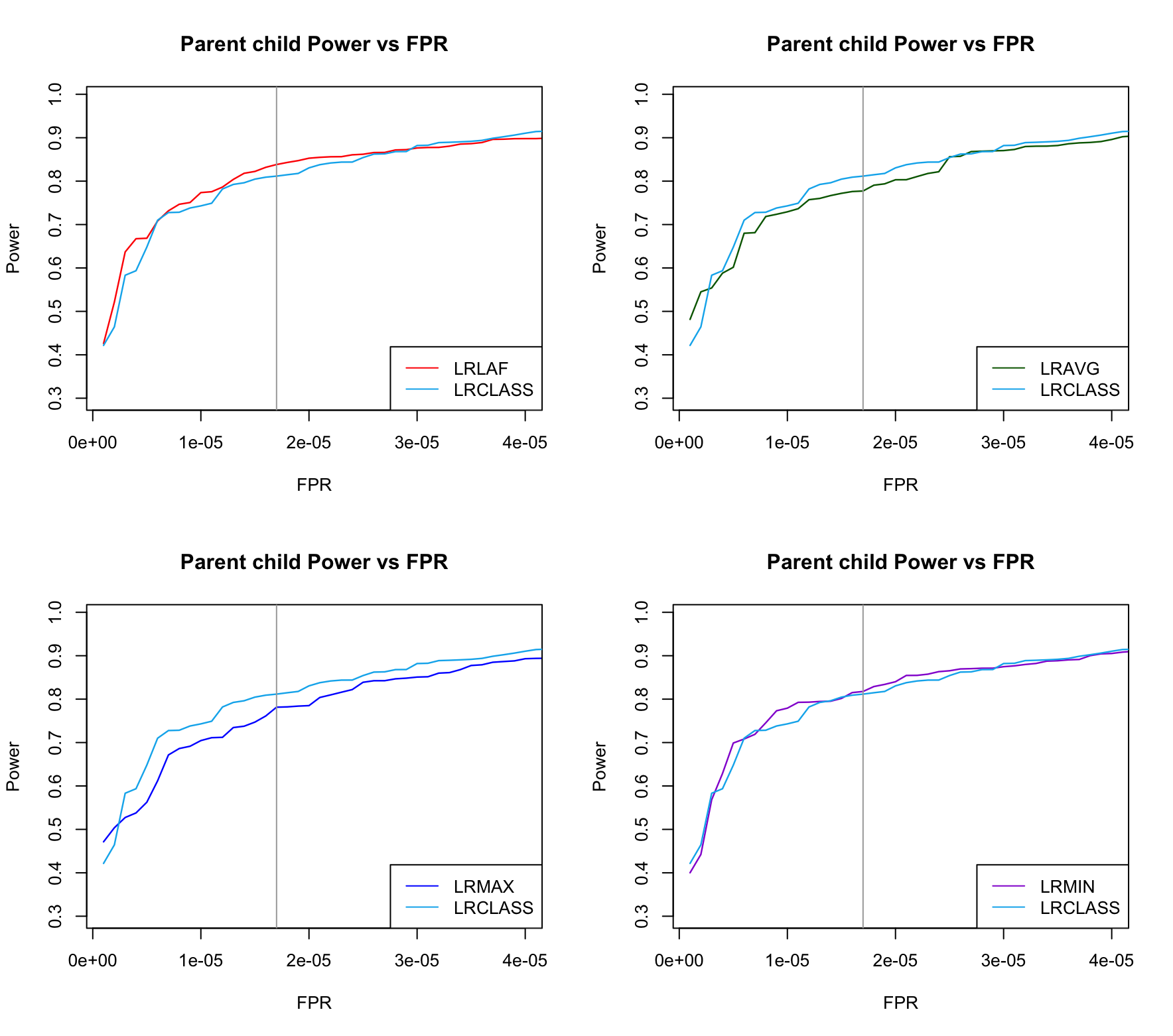}
\caption{Power comparison of LRCLASS to existing methods for parent-child testing}\label{pcclass}
\end{figure}

\begin{figure}[t]
\centering
\includegraphics[height=12cm,keepaspectratio=true]{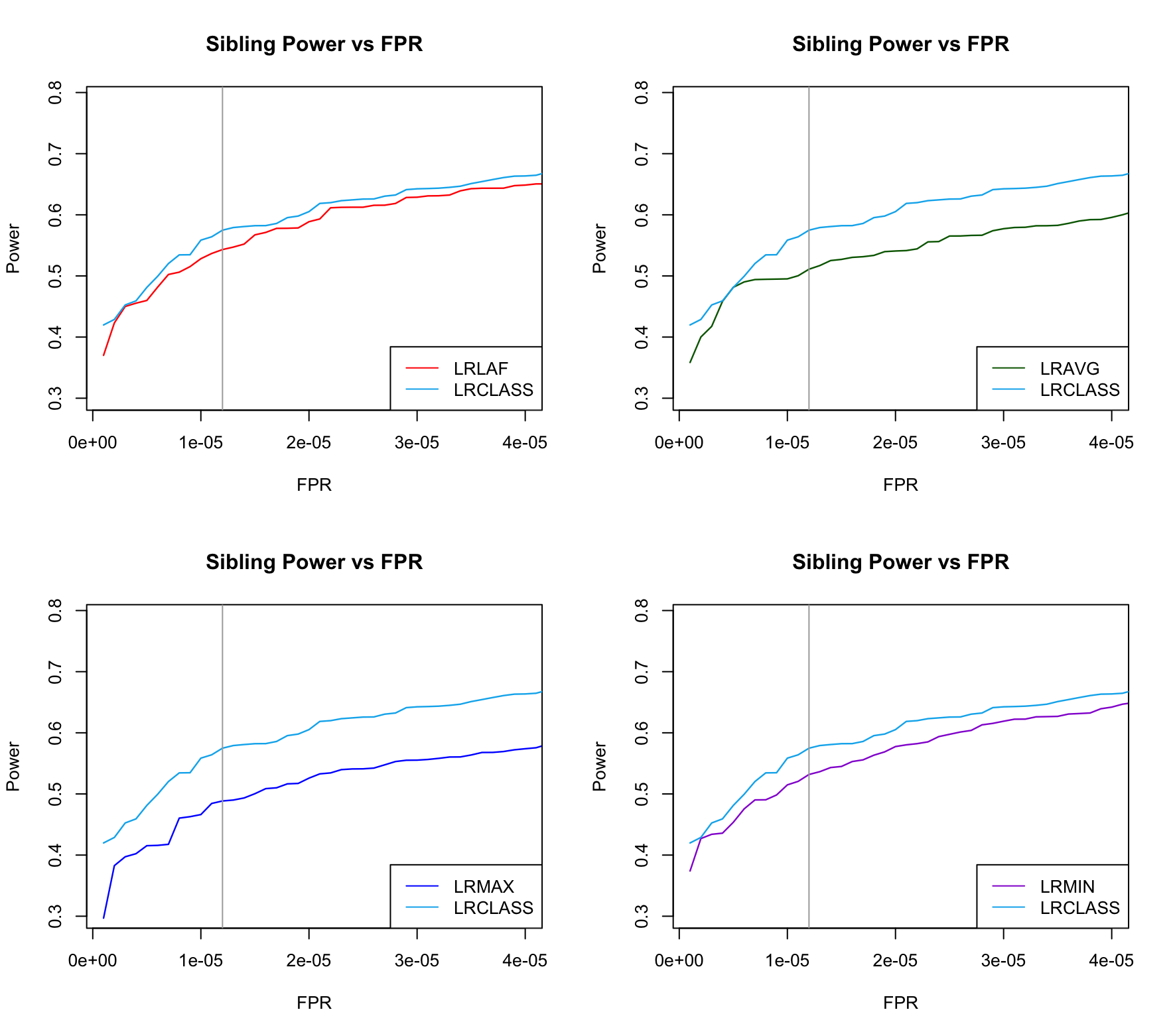}
\caption{Power comparison of LRCLASS to existing methods for full-sibling testing}\label{sbclass}
\end{figure}

To further investigate the performance of LRCLASS, power comparisons were visualized by plotting its power against each of the existing methods. The analysis was extended to include a range of FPRs, spanning from $10^{-6}$ to $4 \times 10^{-5}$. Figures \ref{pcclass} and \ref{sbclass} depict the power curves for parent-child and full-sibling relationship testing, respectively, as functions of the FPR. In these figures, power is displayed on the y-axis, with the statistic corresponding to the upper curve demonstrating greater power than that associated with the lower curve. Additionally, the vertical lines in Figures \ref{pcclass} and \ref{sbclass} represent the suggested FPRs of $1.7 \times 10^{-5}$ and $1.2 \times 10^{-5},$ respectively. 

From Figure \ref{pcclass}, the results indicate that LRCLASS consistently outperforms LRMAX across nearly the entire range of FPRs considered. However, its performance relative to LRLAF, LRAVG, and LRMIN is less conclusive, as no clear dominance is observed throughout the FPR range analyzed. For the full-sibling relationship test, as shown in Figure \ref{sbclass}, LRCLASS demonstrates superior performance compared to all other statistics across the entire range of FPRs considered.

\subsection{Classification Comparison}

Tables \ref{tab: eval gen} and \ref{tab: eval real} present the performance evaluation of each classification method using the validation dataset and the test dataset, respectively. The metrics we used include {micro- (Mic.) and macro- (Mac.)} averaged accuracy, precision, recall, specificity, error rate, and F1-score, along with Cohen's Kappa and Matthews Correlation Coefficient (MCC). The best-performing method for each criterion is emphasized in bold.

\begin{table}[t]
\footnotesize
\centering
\caption{Classification performance of each method on the validation dataset} \label{tab: eval gen}
\tabcolsep=20.5pt
\begin{tabular}{ccccc}
\hline \hline
\multicolumn{2}{c}{Metrics}& Naive Bayes & Logistic Mthd A & Logistic Mthd B \\\hline
Accuracy     & Mac. & 0.8232 & \textbf{0.8259} & 0.8237 \\
             & Mic. & 0.8469 & \textbf{0.8493} & 0.8472 \\ \hline
Precision    & Mac. & 0.6916 & \textbf{0.6975} & 0.6919 \\
             & Mic. & 0.6938 & \textbf{0.6985} & 0.6944 \\ \hline
Recall       & Mac. & 0.6938 & \textbf{0.6985} & 0.6944 \\
             & Mic. & 0.6938 & \textbf{0.6985} & 0.6944 \\ \hline
Specificity  & Mac. & \textbf{0.8511} & 0.8500 & 0.8488 \\
             & Mic. & 0.8979 & \textbf{0.8995} & 0.8981 \\ \hline
Error        & Mac. & 0.1768 & \textbf{0.1741} & 0.1763 \\
             & Mic. & 0.1531 & \textbf{0.1507} & 0.1528 \\ \hline
F1-score     & Mac. & 0.6857 & \textbf{0.6864} & 0.6824 \\
             & Mic. & 0.6938 & \textbf{0.6985} & 0.6944 \\ \hline
\multicolumn{2}{c}{Cohen's Kappa} & 0.5504 & \textbf{0.5548} & 0.5490 \\ \hline
\multicolumn{2}{c}{MCC}           & 0.5537 & \textbf{0.5605} & 0.5542 \\
\hline \hline
\end{tabular}
\end{table}

\begin{table}[t]
\footnotesize
\centering
\caption{Classification performance of each method on the test dataset} \label{tab: eval real}
\tabcolsep=20.5pt
\begin{tabular}{ccccc}
\hline \hline
\multicolumn{2}{c}{Metrics}& Naive Bayes & Logistic Mthd A & Logistic Mthd B \\\hline
Accuracy     & Mac. & \textbf{0.7237} & 0.7219 & 0.7197 \\
             & Mic. & \textbf{0.7346} & 0.7311 & 0.7292 \\ \hline
Precision    & Mac. & 0.4924 & \textbf{0.5003} & 0.4928 \\
             & Mic. & \textbf{0.4693} & 0.4622 & 0.4584 \\ \hline
Recall       & Mac. & \textbf{0.4693} & 0.4622 & 0.4584 \\
             & Mic. & \textbf{0.4693} & 0.4622 & 0.4584 \\ \hline
Specificity  & Mac. & 0.7869 & \textbf{0.7915} & 0.7896 \\
             & Mic. & \textbf{0.8231} & 0.8207 & 0.8195 \\ \hline
Error        & Mac. & \textbf{0.2763} & 0.2781 & 0.2803 \\
             & Mic. & \textbf{0.2654} & 0.2689 & 0.2708 \\ \hline
F1-score     & Mac. & \textbf{0.4462} & 0.4435 & 0.4378 \\
             & Mic. & \textbf{0.4693} & 0.4622 & 0.4584 \\ \hline
\multicolumn{2}{c}{Cohen's Kappa} & \textbf{0.2607} & 0.2566 & 0.2508 \\ \hline
\multicolumn{2}{c}{MCC}           & \textbf{0.2765} & 0.2693 & 0.2636 \\
\hline \hline
\end{tabular}
\end{table}

As presented in Table \ref{tab: eval gen}, logistic regression with Method A demonstrates the best classification performance on the validation set across all evaluated criteria, except for specificity. Conversely, when classification is applied to the test dataset, as shown in Table \ref{tab: eval real}, the Naive Bayes classifier outperforms in terms of recall, accuracy, error, and F1-score, while logistic regression with Method A performs best when evaluated using precision and specificity. It is also observed that the performance of logistic regression with Method B is slightly lower than that of Method A.

Lastly, Table \ref{tab: conf mat real} presents the confusion matrices for the three classification methods applied to the test dataset, where predicted class labels are compared against the actual (true) labels. The models categorize data points into one of four subpopulations: NO, NE, CT, or SO. Each diagonal entry in these matrices represents a correct classification, where the true and predicted subpopulations align. 

\begin{table}[t]
\footnotesize
\centering
\caption{Confusion matrices for the three classification models on the test dataset} \label{tab: conf mat real}
\tabcolsep=7.5pt
\begin{tabular}{ccccccccccccccc}
\hline \hline
\multirow{3}*{} & \multicolumn{4}{c}{Logistic Mthd A} & & \multicolumn{4}{c}{Logistic Mthd B} & & \multicolumn{4}{c}{Naive Bayes} \\\cline{2-15}
 & \multicolumn{4}{c}{Truth} & & \multicolumn{4}{c}{Truth} & & \multicolumn{4}{c}{Truth} \\\cline{2-5} \cline{7-10} \cline{12-15}
 & NO & NE & CT & SO & & NO & NE & CT & SO & & NO & NE & CT & SO \\\hline
Pred. NO & 54 & 9   & 19 & 9  & & 50 & 10  & 20  & 11 & & 59 & 12  & 23 & 12 \\
Pred. NE & 84 & 219 & 75 & 84 & & 87 & 218 & 75  & 84 & & 90 & 239 & 90 & 96 \\
Pred. CT & 54 & 65  & 99 & 61 & & 54 & 65  & 101 & 59 & & 40 & 44  & 78 & 43 \\
Pred. SO & 9  & 10  & 19 & 56 & & 10 & 10  & 16  & 55 & & 13 & 9   & 20 & 59 \\
\hline
Total    & 201& 303 & 212& 210& & 201& 303 & 212 & 209& & 202& 304 & 211& 210 \\
\hline \hline
\end{tabular}
\end{table}

The overall classification accuracy is calculated by dividing the sum of the diagonal entries by the total number of observations in each confusion matrix. For logistic regression using Method A, the estimated correct classification rate is
$(54 + 219 + 99 + 56)/926 = 46.22\%$. For logistic regression using Method B, the estimated rate is 
$(50 + 218 + 101 + 55)/925 = 45.84\%$.
Finally, the estimated correct classification rate for Naive Bayes classification is
$(59 + 239 + 78 + 59)/927 = 46.93\%$.
Note that the original test dataset contained a total of 929 observations. However, due to technical constraints, some observations were excluded for specific methods. For logistic regression, exclusions occurred because certain alleles in the test dataset had never been observed in the training dataset, rendering the method inapplicable. In the case of Naive Bayes, the constraint arose from misalignment between the allele frequency data and the raw data. Specifically, while the allele frequency dataset included the probability of allele 34 at locus D21S11, the test dataset recorded this information as 34.1 or 34.2, creating ambiguity. As a result, such uncertain cases were excluded from the analysis.

\section{Discussions}

In terms of power comparison, as shown in Table \ref{powertable}, the performance can be categorized into two distinct groups. The first group includes LRLAF, LRMIN, and LRCLASS, all of which demonstrate strong performance in both the parent-child and full-sibling tests. The second group consists of LRAVG and LRMAX, which clearly exhibit lower power compared to the other statistics. This observation is consistent with the results of previous studies conducted on the Colorado population (Kooakachai \emph{et al.}, 2019). In particular, we recommend using the proposed statistic LRCLASS for the full-sibling test in the Thai population, as it consistently outperforms all other statistics, as illustrated in Figure \ref{sbclass}. There are no significant limitations associated with using these test statistics, as the computational time required for each method is comparable.

From Tables \ref{tab: eval gen} and \ref{tab: eval real}, it is evident that the three classification methods demonstrate comparable performance in classifying DNA profiles. As presented in Table \ref{tab: eval real}, the Naive Bayes classifier demonstrates superior performance across recall, accuracy, error, and F1-score, underscoring its suitability for tasks demanding well-rounded effectiveness across multiple metrics. In contrast, logistic regression with Method A excels in precision and specificity, making it particularly advantageous in contexts where minimizing false positive classifications is critical, such as scenarios where the cost or risk of false positives outweighs that of false negatives. Logistic regression with Method B exhibits consistent performance but slightly underperforms compared to both Naive Bayes and logistic regression with Method A across most evaluation metrics.

When evaluating the correct classification rate, the Naive Bayes classifier achieves the highest estimated rate at 46.93\%, with a 95\% confidence interval of (0.437, 0.502). However, this difference is not statistically significant when compared to the other methods. Logistic regression using Method A demonstrates an estimated correct classification rate of 46.22\%, accompanied by a 95\% confidence interval of (0.430, 0.495), which overlaps substantially with the performance of the Naive Bayes classifier. Overall, these findings suggest that the three classification methods exhibit comparable performance, with no single method showing a distinct advantage over the others.

As presented in Table \ref{tab: conf mat real}, all classification methods achieve the highest accuracy in identifying DNA profiles from the NE class. Among them, the Naive Bayes classifier performs the best, correctly identifying 239 out of 304 NE samples, equating to approximately 78.6\%. In contrast, the models struggle significantly with other classes, where the accuracy of correct classifications remains below 50\% across all methods. This suggests that for individuals from the Thai population, DNA profiles from the northeastern region are the most consistently classified using both logistic regression and the Naive Bayes method, whereas profiles from other regions show considerably lower classification accuracy.

To discuss the pros and cons of each method, we first note that Naive Bayes is advantageous due to its simplicity and faster computation, as it does not require a training dataset, unlike logistic regression, which necessitates a training phase before classification can be applied. The Naive Bayes method benefits from reduced computational time; however, logistic regression may be preferable when working with specific subpopulations. For instance, as shown in Table \ref{tab: conf mat real}, logistic regression performs better in classifying DNA profiles from the central region of Thailand. Specifically, Methods A and B of logistic regression correctly classified 99 and 101 samples, respectively, out of 212, achieving an accuracy of approximately 46–47\%. In contrast, Naive Bayes yielded a lower accuracy of 36\%, correctly classifying only 78 out of 211 samples. This discrepancy highlights the bias of the Naive Bayes method, which performs well for the Thai Northeastern subgroup but not for the Central subgroup.

Finally, among the two forms of logistic regression proposed, we recommend Method A, which uses 30 predictors by treating each allele across 15 loci as a separate predictor. This approach demonstrates superior overall performance, as evidenced by the results presented in Tables \ref{tab: eval gen} and \ref{tab: eval real}.

\section{Conclusions}

In this study, we introduced a novel approach to familial DNA testing, specifically designed for evaluating parent-child and full-sibling relationships with a population substructure. The proposed method LRCLASS employs a two-step process: classification is performed first, and the resulting classifications are subsequently used to compute the likelihood ratio. Our findings highlight the exceptional performance of the LRCLASS statistic, particularly when paired with the Naive Bayes classifier, which demonstrates superior power in full-sibling relationship testing within the Thai population compared to all existing statistics. Additionally, logistic regression methods were investigated as alternative classification tools, with their performance shown to be comparable to that of the Naive Bayes classifier.

We propose several potential avenues for future research to build upon the findings of this study. First, while the LRCLASS, LRLAF, and LRMIN statistics demonstrate strong performance, it remains an open question whether they represent the most powerful approaches for relatedness testing. Identifying a new statistic or even more complex tools for classification that consistently outperform existing ones could be transformative. Since there is currently no theoretical framework for comparing the efficacy of these statistics, further methodological advancements and theoretical developments in this area would be invaluable.

Second, although our hypothesis tests primarily focused on parent-child and full-sibling relationships, future research could expand to include more complex familial relationships, such as half-siblings, first cousins, second cousins, and double first cousins. Examining how statistical power varies with the complexity of the relationship could provide deeper insights into the capabilities and limitations of current methods.

Finally, our analysis was conducted under the assumption of the Thai population with its distinct regional substructure. Extending this analysis to other populations with varying substructures would be a meaningful step, enabling cross-population comparisons and shedding light on how population-specific genetic diversity impacts the performance of relatedness testing statistics. Such work could contribute to a broader understanding of the generalizability and adaptability of these methods across diverse genetic landscapes.

\section*{Acknowledgement}
The authors wish to express their gratitude to the editor, associate editor, and referees for their valuable comments and suggestions, which significantly enhanced the quality of this article.

\section*{Ethics approval and consent to participate}
The dataset used in this study was approved by the ethics committees of the Faculty of Medicine, Ramathibodi Hospital, Mahidol University, Thailand (COA No. MURA2023/899). The requirement for informed consent was waived for each participant.

\section*{Availability of data and material}
The training datasets generated during the current study, along with the test dataset, are available from the corresponding author upon reasonable request. The allele frequency distributions, which serve as the basis for the generated datasets, can be found in the public literature (Shotivaranon \emph{et al.}, 2009).

\section*{Authors' contributions}
AJ Conceptualization, methodology, software, formal analysis for the classification comparison part, investigation, writing – original draft.
CL Conceptualization, methodology, software, formal analysis for the power comparison part, investigation. 
BR Conceptualization, methodology, dataset provision.
JS Conceptualization, methodology, dataset provision. 
MK Conceptualization, methodology, software, writing – review and editing, supervision, project administration.
All authors read and approved the final manuscript.

\end{document}